\documentclass[9pt,journal]{IEEEtran}
\ifCLASSOPTIONcompsoc
\usepackage[nocompress]{cite}
\else
\usepackage{cite}
\fi
\usepackage{graphicx}
\usepackage{float}
\usepackage{amssymb}
\usepackage{amsmath}
\usepackage{dsfont}
\usepackage{amsthm}
\usepackage{graphicx}
\usepackage{mathrsfs,doi,color}
\usepackage{subfigure}
\usepackage{epstopdf}
\usepackage{hyperref}

\newtheoremstyle{mythm}{1.5ex plus 1ex minus .2ex}{1.5ex plus 1ex minus .2ex} {\rm}{\parindent}{\it\it}{\rm{:}}{1em}{}
\theoremstyle{mythm}

\renewcommand{\IEEEQED}{\IEEEQEDopen}

\newtheorem{theorem}{Theorem}[section]
\newtheorem{corollary}{Corollary}[section]
\newtheorem{lemma}{Lemma}[section]
\newtheorem{proposition}{Proposition}[section]
\newtheorem{definition}{Definition}[section]
\newtheorem{remark}{Remark}

\ifCLASSINFOpdf
\else
\fi
\begin{document}

\title{Convergence Analysis of Weighted-Median Opinion Dynamics with Prejudice}

\author{ Ruichang~Zhang,  Zhixin~Liu,~\IEEEmembership{Member,~IEEE}, Ge~Chen,~\IEEEmembership{Senior Member,~IEEE},  Wenjun~Mei,~\IEEEmembership{ Member,~IEEE}
	
	\IEEEcompsocitemizethanks{\IEEEcompsocthanksitem This work was supported by Natural Science Foundation
		of China under Grant T2293772, and the Strategic Priority Research
		Program of Chinese Academy of Sciences under Grant No.
		XDA27000001.
		
		\IEEEcompsocthanksitem
	   Ruichang Zhang, Zhixin Liu, Ge Chen are with the Key Laboratory of Systems and
		Control, Academy of Mathematics and Systems Science, Chinese Academy of Sciences, Beijing 100190,
		China, and also with the School of
		Mathematical Sciences, University of Chinese Academy of Sciences,
		Beijing 101408, China (e-mail: zrc20@amss.ac.cn, lzx@amss.ac.cn, chenge@amss.ac.cn).
		Wenjun Mei is with the Department of Mechanics and Engineering Science, Peking University (e-mail: mei@pku.edu.cn).
	}
}

\IEEEtitleabstractindextext{%
  \begin{abstract}

The Friedkin-Johnsen (FJ) model introduces prejudice into the opinion evolution and has been successfully validated in many practical scenarios; however, due to its weighted average mechanism, only one prejudiced agent can always guide all unprejudiced agents synchronizing to its prejudice under the connected influence network, which may not be in line with some social realities. To fundamentally address the limitation of the weighted average mechanism, a weighted-median opinion dynamics has been recently proposed; however, its theoretical analysis is challenging due to its nonlinear nature. This paper studies the weighted-median opinion dynamics with prejudice, and obtains the convergence and convergence rate when all agents have prejudice, and a necessary and sufficient condition for asymptotic consensus when a portion of agents have prejudice. These results are the
first time to analyze the  discrete-time and synchronous opinion dynamics with the weighted median mechanism, and address the phenomenon of the FJ model that connectivity leads to consensus when a few agents with the same prejudice join in an unprejudiced group.


\end{abstract}

}

\maketitle

\IEEEdisplaynontitleabstractindextext

\IEEEpeerreviewmaketitle

\renewcommand{\thesection}{\Roman{section}}
\section{Introduction}
\renewcommand{\thesection}{\arabic{section}}

Opinion dynamics play an important role in describing the generation and evolution of social opinions, which has  extensive research interests from various  fields \cite{becker2017network,galesic2019statistical,nowak2019nonlinear,dandekar2013biased}. A seminal model on opinion dynamics called the DeGroot model \cite{degroot1974reaching}  considers the interaction between agents over a directed weighted graph known as the influence network, in which the opinions of agents  are updated through the weighted average of their neighbors' opinions. This simple linear mechanism enables the strong connectivity of the influence network to ensure asymptotic consensus of opinions; however, a widely observed phenomenon is that although people tend to become more alike when they interact with each other, differences of opinions are widely present in connected social networks \cite{abelson1964mathematical,axelrod1997dissemination}. Thus, the DeGroot model does not fully reflect the flexibility shown by natural groups in opinion evolution \cite{9736598}.

To overcome the above limitation of the DeGroot model, a series of new models have been proposed to more objectively describe the law of agent opinion change, like the well-known Friedkin-Johnsen (FJ)  model \cite{friedkin1990social},
bounded-confidence model \cite{deffuant2001mixing,rainer2002opinion} and Altafini model \cite{altafini2012consensus}, which have achieved many beautiful results    \cite{frasca2013gossips,parsegov2016novel,bindel2015bad,ghaderi2014opinion,ishii2010distributed,proskurnikov2017opinion,etesami2015game,chazelle2016inertial,chen2020convergence}. However, most of the existing research is based on the weighted average mechanism which possesses the nature that the greater the distance between agent opinions, the stronger the attraction \cite{mei2022micro,9736598}, and this nature may cause some  problems. For example, the FJ  model \cite{friedkin1990social} assumes that each agent has a prejudice, and incorporates its prejudice into the  weighted average  opinion evolution. Despite the FJ model has been validated in many practical scenarios \cite{childress2012cultural,friedkin2016network}, however,
when a prejudiced agent joins in
an unprejudiced group, if the influence network is connected then
all agents’ opinions will synchronize to the prejudice of the
prejudiced agent under the FJ model \cite{proskurnikov2017opinion}. This phenomenon should not be in line with some social realities \cite{abelson1964mathematical,axelrod1997dissemination}.

In order to fundamentally resolve the limitation of the weighted average mechanism, a weighted-median opinion dynamics has been proposed \cite{mei2022micro} recently, in which the opinions of agents  are updated through the weighted median of their neighbors' opinions. Compared to the weighted average mechanism, the weighted-median mechanism can better explain  the diversity of opinions in real social groups, which has once puzzled many researchers \cite{abelson1964mathematical,axelrod1997dissemination}.
Also, experiments and online data have validated the effectiveness of the weighted-median mechanism in predicting opinion evolution and depicting the characteristics of opinion evolution in real-world scenarios \cite{mei2022micro}. For example, the distribution of public opinion observed from experience indicates that as the group size or clustering coefficient increases, the likelihood of the group reaching consensus gradually decreases \cite{mei2022micro}. Despite its many advantages, the weighted-median model is nonlinear and its theoretical analysis is difficult, and  its convergence has been proved only for the time asynchronous case \cite{mei2022convergence}.

%
%

Motivated by the above observations, this paper investigates a model called \emph{the weighted-median opinion dynamics with prejudice}.
In this model the update of each agent's opinion is based on the collective impact of the prejudice and the weighted-median  mechanism, which takes the advantages of both the weighted-median and the FJ model. An agent is called prejudiced if it is affected by the prejudice in each update step, otherwise it is called unprejudiced.
The main contributions of this paper can be summarized as follows.
\begin{itemize}
	\item We analyze the discrete-time and synchronous opinion dynamics with the weighted-median  mechanism for the first time. In details,
when all agents have prejudice, we provide the convergence and the negative exponential convergence rate of the  weighted-median opinion dynamics with prejudice;
 when a portion of agents have prejudice,  we give a necessary and sufficient condition for asymptotic consensus, which depends on whether the influence network exists a cohesive set consisting of unprejudiced agents only.
 \item Our results show that when a few agents with the same prejudice join in  an unprejudiced and connected group, the asymptotic consensus may not necessarily be reached,
 which is different from the FJ model. From this respect, our system may be more realistic than the FJ model, especially in the large-scale social networks.
\end{itemize}

The paper is organized as follows. Section \ref{Mod_sec} introduces  basic definition and related model. Section \ref{WPOD} and Section \ref{MWPOD} introduce the weighted-median  opinion dynamics with prejudice and analyze the convergence behavior. Section \ref{NE3} presents numerical simulation and Section \ref{CR}  concludes the paper.

\renewcommand{\thesection}{\Roman{section}}
\section{Basic Definition and Related Work}\label{Mod_sec}
\renewcommand{\thesection}{\arabic{section}}


This paper considers a group of $n$ agents.  Let $\mathcal{V}:=\left\{1,  2,  \ldots,  n\right\}$ be the index set of all agents.
Each agent $i\in\mathcal{V}$ has an opinion $x_i(t)\in\mathbb{R}$ at each time instant $t\in\mathbb{N}$,  and set $\boldsymbol{x}(t):=(x_1(t),  \ldots,  x_n(t))^{\top}$.
We assume the evolution of  each agent's opinion is affected by its neighbors, and the neighborhood relationships between agents is represented by a directed and weighted \emph{influence network}.
Denote by $\mathcal{G}(W)$ the influence network  associated
with the weighted adjacency matrix $W=(w_{ij})_{n\times n}$, where  $w_{ij}\in [0,1]$ represents how much the agent $i$ is influenced by  $j$.
We  call $W$ the \emph{influence matrix}. Throughout this paper we assume the non-negative influence matrix $W$ is  row-stochastic, i.e.,  $W \boldsymbol{1}_n =\boldsymbol{1}_n$,  where
$\boldsymbol{1}_n=(1,\ldots,1)^{\top}$ is an $n$-dimensional vector.

\subsection{Weighted-median opinion dynamics}

Let $\boldsymbol{\vartheta}=(\vartheta_1,  \ldots,  \vartheta_n)^{\top}\in [0,1]^n$ be a weight vector satisfying $\sum _{i=1}^{n}\vartheta_i=1$.
We introduce the definition of weighted median first.
\begin{definition}[Weighted median]\label{defWM}
For a  vector $\boldsymbol{x}=(x_1,  \ldots,  x_n)^{\top}\in\mathbb{R}^n$, we say ${x}^{\ast }\in \cup_{i=1}^n\{x_i\}$ is a \emph{weighted median} of $\boldsymbol{x}$ associated with the weight vector $\boldsymbol{\vartheta}$ if
\begin{equation*}
	\sum _{i:x_i<{x}^{\ast }}\vartheta_i\leq\dfrac{1}{2} ~~\mbox{and}~~
	\sum _{i:x_i>{x}^{\ast }}\vartheta_i\leq\dfrac{1}{2}.
\end{equation*}
\end{definition}
The following lemma tells us the existence of  weighted median and under what conditions the weighted median is unique.
\begin{lemma}[Appendix A in \cite{mei2022micro}]\label{lemma00}
If there exists ${x}^{\ast }\in\cup_{i=1}^n\{x_i\}$ such that
$$ \sum _{i:x_i<{x}^{\ast }}\vartheta_i<\dfrac{1}{2},~~ \sum _{i:x_i={x}^{\ast }}\vartheta_i>0,~~\mbox{and}~~\sum _{i:x_i> {x}^{\ast }}\vartheta_i<\dfrac{1}{2},$$
then ${x}^{\ast }$  is the   unique weighted median of  $\boldsymbol{x}$   associated with $\boldsymbol{\vartheta}$;
Otherwise there exists $z\in\cup_{i=1}^n\{x_i\}$ such that
$$\sum _{i:x_i<z}\vartheta_i=\sum _{i:x_i\geq z}\vartheta_i=\dfrac{1}{2},$$
then the weighted median of $\boldsymbol{x}$ associated with $\boldsymbol{\vartheta}$ is not unique.
\end{lemma}

Next we introduce the \emph{weighted-median opinion dynamics} first proposed in \cite{mei2022micro}.
Consider a group of $n$ agents in an influence network $\mathcal{G}(W)$ associated with a row-stochastic influence matrix $W$. The opinions of all agents are updated according to the following equation:
\begin{equation}\label{Meidynamic}
x_i(t+1)=\mbox{Med}_i(\boldsymbol{x}(t);W),~~\forall i\in\mathcal{V}, t\in\mathbb{N},
\end{equation}
where $\mbox{Med}_i(\boldsymbol{x};W)$ is the weighted median of $\boldsymbol{x}$ associated with the weight vector $(w_{i1},  \ldots,  w_{in})^{\top}$. If the weighted median of $\boldsymbol{x}$ associated with the weight vector $(w_{i1},  \ldots,  w_{in})^{\top}$ is not unique, we take $\mbox{Med}_i(\boldsymbol{x},W)$ as the weighted median closest to $x_i$.


 Previous work \cite{mei2022convergence} characterizes the set composed of all equilibrium points and establishes the convergence analysis of asynchronous weighted-median opinion dynamics.
 However, the convergence condition is much complex  when all agents update their opinions at the same time, and its theoretical proof remains unresolved.

\subsection{Friedkin-Johnsen model}

In FJ model \cite{friedkin1990social}, each agent has persistent attachment to its prejudice and also influenced by other agents.
Let $\Lambda:=\mbox{diag}(\lambda_1,  \ldots,  \lambda_n)$ be agents' susceptibilities to the social influence with $\lambda_i\in[0,1]$ for all $i\in\mathcal{V}$, and $\boldsymbol{u}:=(u_1,  \ldots,  u_n)^{\top}\in\mathbb{R}^n$ be the prejudices of the agents. Usually
$\boldsymbol{u}$ is set to be $\boldsymbol{x}(0)$.
The update equation of opinions in the FJ model can be formulated by
\begin{equation}\label{FJ}
\boldsymbol{x}(t+1)=\Lambda \boldsymbol{u}+(I_n-\Lambda)W\boldsymbol{x}(t),~~t\in\mathbb{N},
\end{equation}
where $I_n$ is an $n$-order identity matrix.

The FJ model has attracted much attention of researches. For examples, a sufficient condition for the stability of the FJ model is introduced in \cite{frasca2013gossips}, and the convergence condition of the FJ model is given in \cite{parsegov2016novel}.
The FJ model also has elegant game theory \cite{bindel2015bad} and electrical \cite{ghaderi2014opinion} interpretations.

\renewcommand{\thesection}{\Roman{section}}
\section{Weighted-median Prejudice Opinion Dynamics}\label{WPOD}
\renewcommand{\thesection}{\arabic{section}}

In this paper, we study the \emph{weighted-median  opinion dynamics with prejudice} which adopts the advantages of the weighted-median  mechanism in the weighted-median opinion dynamics (\ref{Meidynamic}) and  persistent prejudice attachment in the FJ model (\ref{FJ}).
We still consider a group of $n$ agents in an influence network $\mathcal{G}(W)$ associated with an influence matrix $W$. At each time $t$ every agent $i$ updates its opinion $x_i(t)\in\mathbb{R}$ synchronously by
\begin{align}\label{doc10}
	x_i(t+1)=\lambda_i u_i+(1-\lambda_i)\mbox{Med}_i(\boldsymbol{x}(t);  W), ~\forall i\in\mathcal{V}, t\in\mathbb{N}.
\end{align}
In system (\ref{doc10}), we call agent $i$ \emph{prejudiced} if $\lambda_i\in(0,  1]$, i.e., agent $i$ is
influenced by the prejudice $u_i$ at each time instant $t$; otherwise,  we call agent $i$ \emph{unprejudiced},  i.e., the agent $i$ is not influenced by the prejudice at all the time  ($\lambda_i=0$). We also call the system (\ref{doc10}) as the  \emph{weighted-median prejudice opinion dynamics} if
all agents have prejudices, and as the \emph{weighted-median opinion dynamics with partial  prejudice} if
the system is mixed by prejudiced and unprejudiced  agents.

This section discusses the theoretical properties of the weighted-median prejudice opinion dynamics.
Denote $$\mbox{Med}(\boldsymbol{x}(t);W):=(\mbox{Med}_1(\boldsymbol{x}(t);W),  \ldots,  \mbox{Med}_n(\boldsymbol{x}(t);W))^{\top}.$$
Then the system (\ref{doc10}) can be rewritten as
\begin{align}\label{doc0}
	\boldsymbol{x}(t+1)=\Lambda \boldsymbol{u}+(I_n-\Lambda)\mbox{Med}(\boldsymbol{x}(t);W),~~\forall t\in\mathbb{N}.
\end{align}
In our system, the agent $i$ evaluate the two factors $u_i$ and $\mbox{Med}_i(\boldsymbol{x}(t);  W)$ based on susceptibility $\lambda_i$  to hold the opinion at the next time.

\begin{definition}
If for any $\boldsymbol{x}(0)\in \mathbb{R}^n$  there exists a constant $x^{\ast}\in\mathbb{R}$ such that $\lim_{t\rightarrow\infty}x_i(t)=x^{\ast},\forall i\in\mathcal{V}$,  then we say the system (\ref{doc0})  achieves consensus
asymptotically.
\end{definition}

Set $\lambda_{\min}:=\min\left\{\lambda_1,\ldots, \lambda_n\right\}$ and $\lambda_{\max}:=\max\left\{\lambda_1,\ldots, \lambda_n\right\}$.
We first give the convergence and convergence rate of the weighted-median prejudice opinion dynamics.

\begin{theorem}[Convergence and convergence rate of weighted-median prejudice opinion dynamics]
	\label{Main_result}
Consider the system  (\ref{doc0}) consists of prejudiced agents only, i.e., $\lambda_i\in(0,  1]$ and $u_i\in\mathbb{R}$ for all $i\in\mathcal{V}$. Then,\\
i) there exists a vector $\boldsymbol{x}^*=({x}^{\ast }_1,\ldots,{x}^{\ast }_n)^{\top}\in\mathbb{R}^n$
depending on the  system parameters $W,\Lambda$ and $\boldsymbol{u}$ only, such that $\lim_{t\rightarrow\infty}\boldsymbol{x}(t)=\boldsymbol{x}^*$ with the convergence rate
$$\|\boldsymbol{x}(t)-\boldsymbol{x}^*\|_{\infty} \leq (1-\lambda_{\min})^t \|\boldsymbol{x}(0)-\boldsymbol{x}^*\|_{\infty},~\forall \boldsymbol{x}(0)\in \mathbb{R}^n, t\in\mathbb{N};$$
ii) the  asymptotic consensus can be achieved
for any initial state if and only if $u_1=u_2=\ldots=u_n$.
\end{theorem}


\begin{remark}
The weighted-median serves as a nonlinear update mechanism, which leads to difficulties in theoretical analysis. For the weighted-median opinion dynamics, the convergence conditions have been established only under the time asynchronous update mechanism, and are still unresolved under the time synchronous update mechanism \cite{mei2022convergence}. However, by combining the  weighted-median model with the FJ model,
 Theorem \ref{Main_result} provides general convergence conditions and a negative exponential convergence rate, which is similar to some widely studied linear dynamics.
\end{remark}
The proof of Theorem \ref{Main_result} uses the well-known Banach's fixed point theorem.
$(\mathbb{R}^n,\|\cdot\|_{\infty})$ is a complete
metric space where $\|\cdot\|_{\infty}$ denotes the infinity norm,  the Banach's fixed point theorem can be formulated as follows.
\begin{lemma}[Banach's Fixed Point Theorem]\label{Banach}
Let $F: \mathbb{R}^n\rightarrow\mathbb{R}^n$ be a contraction on $(\mathbb{R}^n,\|\cdot\|_{\infty})$, i.e., there exists a constant $\alpha\in[0,1)$ such that
$$\|F(\boldsymbol{x})-F(\boldsymbol{y})\|_{\infty} \leq \alpha \|\boldsymbol{x}-\boldsymbol{y}\|_{\infty},~~\forall \boldsymbol{x},\boldsymbol{y}\in \mathbb{R}^n.$$
Then $F(\cdot)$ has a unique fixed point $\boldsymbol{x}^*\in\mathbb{R}^n$, i.e., $F(\boldsymbol{x}^*)=\boldsymbol{x}^*$.
\end{lemma}

Regarding $\mbox{Med}(\boldsymbol{x};W)$ as the weighted median mapping of $\boldsymbol{x}$ from $\mathbb{R}^n$ to  $\mathbb{R}^n$,
we give its non-expansion property as follows.
\begin{lemma}[Non-expansion of weighted median mapping,  Theorem 8 in \cite{han2024continuoustime}]
	\label{lemma1}
	For any $\boldsymbol{x},\boldsymbol{y}\in \mathbb{R}^n$, we have
	$$\|\mbox{Med}(\boldsymbol{x};W)-\mbox{Med}(\boldsymbol{y};W)\|_\infty\leq \|\boldsymbol{x}-\boldsymbol{y}\|_\infty. $$
	
\end{lemma}


\begin{IEEEproof}[Proof of Theorem \ref{Main_result}]
i) Let $$F(\boldsymbol{x}):=\Lambda \boldsymbol{u}+(I_n-\Lambda)\mbox{Med}(\boldsymbol{x};W),~~\forall \boldsymbol{x}\in\mathbb{R}^n.$$
By Lemma \ref{lemma1}, for any $\boldsymbol{x},\boldsymbol{y}\in \mathbb{R}^n$  we get
\begin{eqnarray}\label{mr1proof_1}
\begin{aligned}
&\|F(\boldsymbol{x})-F(\boldsymbol{y})\|_{\infty}\\
&~~=\|(I_n-\Lambda)\mbox{Med}(\boldsymbol{x};W)-(I_n-\Lambda)\mbox{Med}(\boldsymbol{y};W)\|_\infty\\
&~~\leq (1-\lambda_{\min})\|\mbox{Med}(\boldsymbol{x}(t);W)-\mbox{Med}(\boldsymbol{x}(t-1);W)\|_\infty\\
&~~\leq (1-\lambda_{\min})\|\boldsymbol{x}-\boldsymbol{y}\|_\infty,
\end{aligned}
\end{eqnarray}
which means $F(\cdot)$ is a  contraction on $\mathbb{R}^n$. By Lemma \ref{Banach},  $F(\cdot)$ has a unique fixed point $\boldsymbol{x}^*\in\mathbb{R}^n$.
Because $\boldsymbol{x}^*$ is determined by $F(\cdot)$,  $\boldsymbol{x}^*$ depends on the parameters $W,\Lambda$ and $\boldsymbol{u}$ only.

On the other hand, by (\ref{doc0}) we have $\boldsymbol{x}(t+1)=F(\boldsymbol{x}(t))$ for any $t\in\mathbb{N}$. Then
by (\ref{mr1proof_1}) and the fact $F(\boldsymbol{x}^*)=\boldsymbol{x}^*$ we obtain
\begin{eqnarray}
	\begin{aligned}\label{mr1proof_2}
	\|\boldsymbol{x}(t+1)-\boldsymbol{x}^\ast\|_\infty&=\|F(\boldsymbol{x}(t))-F(\boldsymbol{x}^\ast)\|_\infty \\
	&\leq (1-\lambda_{\min})\|\boldsymbol{x}(t)-\boldsymbol{x}^\ast\|_\infty \\
&\leq \ldots \leq (1-\lambda_{\min})^{t+1}\|\boldsymbol{x}(0)-\boldsymbol{x}^*\|_\infty,
	\end{aligned}
\end{eqnarray}
which implies  $\underset{t\to \infty }{\lim}\boldsymbol{x}(t)={\boldsymbol{x}}^{\ast }$.

	ii) $(\Longleftarrow)$ If $u_1=u_2=\ldots=u_n=u^*$, by the definition of $F(\cdot)$ and $\mbox{Med}(\cdot;W)$ we can get
\begin{eqnarray}\label{mr1proof_3}
\begin{aligned}
F(u^*\boldsymbol{1}_n)&=\Lambda u^*\boldsymbol{1}_n+(I_n-\Lambda)\mbox{Med}(u^*\boldsymbol{1}_n;W)\\
&=\Lambda u^*\boldsymbol{1}_n+(I_n-\Lambda)u^*\boldsymbol{1}_n=  u^*\boldsymbol{1}_n.
\end{aligned}
\end{eqnarray}
By i), $F(\cdot)$ has a unique fixed point. From (\ref{mr1proof_3}), we see that  the unique fixed point of the mapping $F(\cdot)$ is  $ u^*\boldsymbol{1}_n $.  Thus, by (\ref{mr1proof_2})  $\underset{t\to \infty }{\lim}\boldsymbol{x}(t)=u^*\boldsymbol{1}_n$ for any initial state, i.e.,  the system  (\ref{doc0})    achieves consensus asymptotically.

 $(\Longrightarrow)$  If the system (\ref{doc0})  asymptotically achieves consensus $u^*$ for any initial state, by i)  $u^*\boldsymbol{1}_n$ is the unique fixed point of the mapping $F(\cdot)$, which means
 \begin{eqnarray}\label{mr1proof_4}
\begin{aligned}
u^*\boldsymbol{1}_n&=F(u^*\boldsymbol{1}_n)=\Lambda \boldsymbol{u}+(I_n-\Lambda)\mbox{Med}(u^*\boldsymbol{1}_n;W)\\
&=\Lambda \boldsymbol{u}+(I_n-\Lambda)u^*\boldsymbol{1}_n.
\end{aligned}
\end{eqnarray}
From (\ref{mr1proof_4}) we have $\Lambda \boldsymbol{u}=\Lambda u^*\boldsymbol{1}_n$. By the
reversibility of the matrix $\Lambda$ we can obtain that $\boldsymbol{u}=u^*\boldsymbol{1}_n$. This complete the proof of the lemma.
\end{IEEEproof}

An unresolved problem for Theorem \ref{Main_result} is the  analytical expression of the limit $\boldsymbol{x}^{\ast}$ for the general system parameters $W,\Lambda$ and $\boldsymbol{u}$.
This problem is difficult to be solved completely. As a substitute we give an explicit expression for $\boldsymbol{x}^{\ast}$  in form.

For any subset $\mathcal{A}\subseteq\mathcal{V}$, define the indicator function
\begin{equation*}
	{\mathbb{I}}_{\mathcal{A}}(x):=
	\left\{
	\begin{aligned}
		&1, ~~~\mbox{if $x\in \mathcal{A}$}\\
		&0, ~~~\mbox{otherwise}
	\end{aligned}\right..
\end{equation*}
\begin{corollary}\label{limitform}
	Consider the system  (\ref{doc0}) with $\lambda_i\in(0,  1]$ and $u_i\in\mathbb{R}$ for all $i\in\mathcal{V}$.  Then there exists a matrix $A\in \mathbb{R}^{n\times n}$ satisfying that in
 each row $i$ of $A$, one entry is $1-\lambda_i$ and all of the other entries are $0$, such that
 the limit point
	\begin{equation} \label{line-matrix}
		{\boldsymbol{x}}^{\ast}=(I_n-A)^{-1}\Lambda{\boldsymbol{u}}.
	\end{equation}
\end{corollary}

\begin{IEEEproof}
	From the proof of Theorem \ref{Main_result}, the limit point ${\boldsymbol{x}}^{\ast}$  of system (\ref{doc0}) is also the unique fixed point of the mapping $F(\boldsymbol{x})=\Lambda \boldsymbol{u}+(I_n-\Lambda)\mbox{Med}(\boldsymbol{x};W)$, i.e., $F(\boldsymbol{x}^*)=\boldsymbol{x}^*$.
	So we have
	\begin{equation}\label{limitdform_1}
		{x}_i^{\ast}=\lambda_i u_i+(1-\lambda_i)\mbox{Med}_i(\boldsymbol{x}^{\ast};W), ~~\forall i\in\mathcal{V}.
	\end{equation}
According to the definition of weighted median, for any $i\in\mathcal{V}$ the value of $\mbox{Med}_i(\boldsymbol{x}^*;W)$ is an entry of the vector $\boldsymbol{x}^*$, which means that there exists an agent $k_i\in \mathcal{V}$ such that $\mbox{Med}_i(\boldsymbol{x}^*;W)={x}_{k_i}^*$. We can rewrite (\ref{limitdform_1}) into the following matrix form.
	\begin{equation}\label{limitdform_2}
		{x}_i^{\ast}=\lambda_i u_i+(1-\lambda_i){x}_{k_i}^*, ~~\forall i\in\mathcal{V}.
	\end{equation}
 Let  $A=(a_{ij})_{n\times n}\in \mathbb{R}^{n\times n}$ be a matrix with $a_{ij}=(1-\lambda_i)\mathbb{I}_{\{k_i\}}(j)$ for all $i,j\in\mathcal{V}$.
Then equations (\ref{limitdform_2}) can be rewritten as
\begin{equation}\label{limitdform_3}
	{\boldsymbol{x}}^{\ast}=\Lambda{\boldsymbol{u}}+A{\boldsymbol{x}}^{\ast}.
\end{equation}
It is clear that $I-A$ is a strictly diagonally dominant matrix. According to the Disk Theorem \cite{horn2012matrix},  $I_n-A$ is
 invertible. Combining this with (\ref{limitdform_3}) yields our result.
\end{IEEEproof}	

\begin{remark}
		In the FJ model (\ref{FJ}), if each agent is prejudiced or affected by prejudiced neighbors, then the analytical expression of  the limit point is $(I_n-(I_n-\Lambda)W)^{-1}\Lambda{\boldsymbol{u}}$  \cite{proskurnikov2017tutorial}. For the weighted-median prejudice opinion dynamics, we can obtain a similar result in the form of limit behavior.

\end{remark}

Although  the matrix $A$ in Corollary \ref{limitform} is  determined by the influence matrix $W$, prejudices $\boldsymbol{u}$ and susceptibilities $\Lambda$,  its exact value is usually difficult to be obtained directly.
However, for some special graph $\mathcal{G}(W)$ like the complete graph, i.e., $w_{ij}=1/n$  for all $i,j\in \mathcal{V}$, the value of matrix $A$  can be calculated.

\begin{proposition}
Consider the system  (\ref{doc0}) with $\lambda_i\in(0,  1]$ and $u_i\in\mathbb{R}$ for all $i\in\mathcal{V}$. Assume that the influence network $\mathcal{G}(W)$ is a complete graph, i.e., $w_{ij}=1/n$ for all $i,j\in\mathcal{V}$. Without loss of generality we assume $u_1\leq \cdots \leq u_n$.
Then,  $\lim_{t\rightarrow\infty}\boldsymbol{x}(t)=(I_n-A)^{-1}\Lambda{\boldsymbol{u}}$ with $A=(a_{ij})_{n\times n}$ satisfying
\begin{equation*}
	a_{ij}=
	\left\{
	\begin{aligned}
		&(1-\lambda_i)\mathbb{I}_{\{\frac{n+1}{2}\}}(j), ~~~~~~~~~~\mbox{if $n$ is odd}\\
		&(1-\lambda_i)\mathbb{I}_{\{\frac{n}{2}+\lfloor\frac{2i-1}{n}\rfloor\}}(j), ~~~\mbox{if $n$ is even}
	\end{aligned}\right..
\end{equation*}
\end{proposition}
\begin{IEEEproof} We first consider the case when  $n$ is even. By Theorem \ref{Main_result}, the limit point  $\boldsymbol{x}^*$ of the system (\ref{doc0}) does not depend on the initial state $\boldsymbol{x}(0)$. For the convenience of analysis, assume the initial value $\boldsymbol{x}(0)$ is equal to $(u_1,  \ldots,  u_n)^{\top}$, which is followed by
\begin{equation*}
	x_i(0)
	\left\{
	\begin{aligned}
		&\leq x_{\frac{n}{2}}(0)=u_{\frac{n}{2}}, ~~~~~~~~~~\mbox{if~} 1\leq i \leq \frac{n}{2} \\
		&\geq x_{\frac{n}{2}+1}(0)=u_{\frac{n}{2}+1}, ~~~~\mbox{if~} \frac{n}{2}+1\leq i \leq n
	\end{aligned}\right..
\end{equation*}	
	Since $w_{ij}=1/n$ for all $i,j\in\mathcal{V}$,  from the definition of weighted median, the value of $\mbox{Med}_i(\boldsymbol{x}(0);W)$ is the element closest to  $x_i(0)$ in the set
$\{x_{\frac{n}{2}}(0)\}\cup\{x_{\frac{n}{2}+1}(0)\}$, which indicates
	\begin{align}\label{prop1_p1}
	\mbox{Med}_i(\boldsymbol{x}(0);W)&=
		\left\{
		\begin{aligned}
			&u_{\frac{n}{2}},  ~~~~~\mbox{if~}1\leq i \leq \frac{n}{2}&\\
			&u_{\frac{n}{2}+1},  ~~\mbox{if~}\frac{n}{2}+1\leq i \leq n&
		\end{aligned}\right..
	\end{align}
 By (\ref{doc10}) and (\ref{prop1_p1}) we have
 \begin{equation}\label{prop1_p2}
	 x_i(1)=\left\{
		\begin{aligned}
			&\lambda_i u_i+(1-\lambda_i)u_{\frac{n}{2}},  ~~~~\mbox{if~}1\leq i \leq \frac{n}{2}&\\
			&\lambda_i u_i+(1-\lambda_i) u_{\frac{n}{2}+1},  ~\mbox{if~}\frac{n}{2}+1\leq i \leq n&
		\end{aligned}\right..
\end{equation}

Now we compute the state $\boldsymbol{x}(2)$.
If $1\leq i \leq \frac{n}{2}$, which means $u_i\leq u_{\frac{n}{2}},$  by
(\ref{prop1_p2}) we have
\begin{equation*}\label{prop1_p3}
	 x_i(1)\leq \lambda_i u_{\frac{n}{2}}+(1-\lambda_i)u_{\frac{n}{2}}=u_{\frac{n}{2}}.
\end{equation*}
Similarly, if $\frac{n}{2}+1\leq i \leq n$ we have  $x_i(1)\geq u_{\frac{n}{2}+1}$. With the similar discussion as that of (\ref{prop1_p1}) and (\ref{prop1_p2}) we have
$\boldsymbol{x}(2)=\boldsymbol{x}(1)$, which means $\boldsymbol{x}(1)$ is the limit point of the system  (\ref{doc0}). Since
\begin{align*}\label{prop1_p4}
	&x_{\frac{n}{2}}(1)~~~= \lambda_{\frac{n}{2}} u_{\frac{n}{2}}+(1-\lambda_{\frac{n}{2}})u_{\frac{n}{2}}=u_{\frac{n}{2}}\\
	&x_{\frac{n}{2}+1}(1)= \lambda_{\frac{n}{2}+1} u_{\frac{n}{2}+1}+(1-\lambda_{\frac{n}{2}+1})u_{\frac{n}{2}}=u_{\frac{n}{2}+1},
\end{align*}
by (\ref{prop1_p2}) we have
\begin{equation*}\label{prop1_p5}
	{x}_i^*=\left\{
	\begin{aligned}
		&\lambda_i u_i+(1-\lambda_i){x}_{\frac{n}{2}}^*,  ~~~~\mbox{if~}1\leq i \leq \frac{n}{2}&\\
		&\lambda_i u_i+(1-\lambda_i) {x}_{\frac{n}{2}+1}^*,  ~\mbox{if~}\frac{n}{2}+1\leq i \leq n&
	\end{aligned}\right.,
\end{equation*}
i.e.,
\begin{equation*}\label{prop1_p6}
{x}_i^*=\lambda_i u_i+(1-\lambda_i){x}_{\frac{n}{2}+\lfloor\frac{2i-1}{n}\rfloor\}}^*.
\end{equation*}
Our result can be obtained by the similar discussion below (\ref{limitdform_2}) in the proof of Corollary \ref{limitform}.

When  $n$ is odd,  similar to (\ref{prop1_p1}), we have
\begin{align*}
	\mbox{Med}_i(\boldsymbol{x}(0);W)&=
	u_{\frac{n+1}{2}},  ~~~~~\forall 1\leq i \leq n.
\end{align*}
Then the result of the proposition can be obtained by the  similar analysis to the case when $n$ is even.
\end{IEEEproof}

\renewcommand{\thesection}{\Roman{section}}
\section{Weighted-median  Opinion Dynamics with Partial Prejudice}\label{MWPOD}
\renewcommand{\thesection}{\arabic{section}}
Recall that agent $i$ is prejudiced if $\lambda_i\in(0,  1]$, and is unprejudiced if $\lambda_i=0$. This section discusses the limit behavior of the system (\ref{doc10}) containing $n_1$ prejudiced agents and $n_2$ unprejudiced agents with
$n_1\geq 1, n_2\geq 1$ and $n_1+n_2=n$. Without loss of generality we assume agents $1,2,\ldots,n_1$ are prejudiced,
and agents $n_1+1,n_1+2,\ldots,n$ are unprejudiced. Let $\mathcal{V}_1:=\{1,2,\ldots,n_1\} \subseteq \mathcal{V}$ be the set of the prejudiced agents and $\mathcal{V}_2:=\{n_1+1,n_1+2,\ldots,n\} \subseteq \mathcal{V}$ be the set of the unprejudiced agents. It is clear that $\mathcal{V}_1 \cap \mathcal{V}_2=\emptyset $ and $\mathcal{V}_1 \cup \mathcal{V}_2=\mathcal{V} $.
Throughout this section, we assume all prejudices $\{u_i\}$ equal $u$. As a result, the dynamics (\ref{doc10}) has the following expression,
\begin{equation}\label{sys2}
	x_i(t+1)=
	\left\{
	\begin{aligned}
		&\lambda_i u+(1-\lambda_i)\mbox{Med}_{i}(\boldsymbol{x}(t);  W),  ~i\in \mathcal{V}_1&\\
		&\mbox{Med}_{i}(\boldsymbol{x}(t);  W),  ~~~~~~~~~~~~~~~~~~~i\in \mathcal{V}_2&
	\end{aligned}\right.
\end{equation}
with $\lambda_i\in(0,1]$ for any $i\in \mathcal{V}_1$.

Before the statement of our main result for opinion dynamics (\ref{sys2}) we need  some
lemmas.
\begin{lemma}\label{lemma0}
Consider a group of $n$ agents in an influence network $\mathcal{G}(W)$ associated with an influence matrix $W$.
If there exists an agent $i \in \mathcal{V}$ and a set $\mathcal{M}  \subset \mathcal{V}$ satisfying
\begin{equation}\label{lemma01a}
	\sum_{j \in \mathcal{M}} w_{ij}
	\left\{
	\begin{aligned}
		&>\frac{1}{2},  ~~~\mbox{if~} i\notin \mathcal{M}\\
		&\geq\frac{1}{2},  ~~~\mbox{if~} i\in \mathcal{M}
	\end{aligned}\right.,
\end{equation}
then
\begin{equation}\label{lemma01}
\underset{j\in \mathcal{M}}{\min}x_j \leq \mbox{Med}_{i}(\boldsymbol{x};  W) \leq \underset{j\in \mathcal{M}}{\max  }x_j,~~\forall \boldsymbol{x}=(x_1,\ldots,x_n)^{\top}\in\mathbb{R}^n.
\end{equation}
\end{lemma}
\begin{IEEEproof}
 Let $x_{k_1}$,  $x_{k_2},  \ldots,  x_{k_n}$ be a re-ordering of $x_1$,  $x_2$,  $\ldots,  x_n$ with
\begin{equation}\label{pl011}
	x_{k_1} \leq x_{k_2} \leq \ldots \leq x_{k_n}.
\end{equation}
Define the index set
\begin{equation}\label{pl012}
J^*:=\Big\{j\in \{1,\ldots,n\}: \sum_{j<j^{\ast}}w_{ik_j} \leq \dfrac{1}{2}, \sum_{j>j^{\ast}}w_{ik_j} \leq \dfrac{1}{2}\Big\}.
\end{equation}
By Lemma \ref{lemma00},  $J^*$ is not an empty set.  Also, according to  Definition \ref{defWM}
we have
\begin{eqnarray}\label{pl012a}
\mbox{Med}_{i}(\boldsymbol{x};  W)\in \bigcup_{j^*\in J^*} \{x_{k_{j^*}}\}.
\end{eqnarray}
Let
\begin{eqnarray}\label{pl012b}
\begin{aligned}
&a:=\min\{i\in \{1,\ldots,n\}: k_i\in \mathcal{M}\},\\
&b:=\max\{i\in \{1,\ldots,n\}: k_i\in \mathcal{M}\}.
\end{aligned}
\end{eqnarray}
Then by (\ref{pl011}) we get
\begin{equation}\label{A}
\mathcal{M} \subseteq \{k_a,k_{a+1},\ldots,k_{b}\}
\end{equation}

 We first consider the case with	$i\notin \mathcal{M}$ . From (\ref{A}) and (\ref{lemma01a}) we have
	\begin{equation*}\label{pl013}
		\sum_{j={a}}^{{b}}w_{ik_j} \geq \sum_{k \in \mathcal{M}} w_{ik} > \dfrac{1}{2},
	\end{equation*}
	 which means
	 	 \begin{equation}\label{pl014}
	 	\sum_{j\geq{a}}w_{ik_j}\geq \dfrac{1}{2},~ \sum_{j\leq{b}}w_{ik_j}\geq \dfrac{1}{2}.
	 \end{equation}
 By (\ref{pl012}) and  (\ref{pl014}) we have ${a} \leq j^{\ast}\leq {b}$ for  $j^*\in J^*$. Then by (\ref{pl011}) we have
\begin{equation}\label{pl015}
 x_{k_{a}} \leq x_{k_{j^*}} \leq x_{k_{b}},~~\forall j^*\in J^*.
 \end{equation}
 Combining this with (\ref{pl012a}) and (\ref{pl012b})  yields our result.
	
We consider the case with $i\in \mathcal{M}$. If $J^*\subset \{a,a+1,\ldots,b\}$, then the inequality (\ref{pl015}) still holds which yields our result.
If there exists $l\in J^*$ but $l\notin \{a,a+1,\ldots,b\}$ such that
$\mbox{Med}_{i}(\boldsymbol{x};  W)=x_{k_l}$.  Without loss of generality we assume that $l<a$. By (\ref{lemma01a}) and (\ref{lemma01}) we have
\begin{eqnarray*}
	\begin{aligned}
		&\sum_{j<a}w_{ik_j}=1- \sum_{j\geq a}w_{ik_j}\leq 1- \sum_{k\in\mathcal{M}}w_{ik} \leq \dfrac{1}{2},\\
		&\sum_{j>a}w_{ik_j} \leq  \sum_{j>l}w_{ik_j} \leq \dfrac{1}{2}.
	\end{aligned}
\end{eqnarray*}
Then by (\ref{pl012}), we have $a\in J^*$  which means  $x_{k_a}$ is  a weighted median of $\boldsymbol{x}$ associated with the weight vector $(w_{i1},w_{i2},\ldots,w_{in})^{\top}$.

On the other hand, since  $i\in \mathcal{M}$ and $l<a$, by (\ref{pl011}) and (\ref{pl012b}) we have $0\leq x_i-x_{k_a}\leq x_i-x_{k_l}$.
Thus, from the assumptions of $x_{k_l}=\mbox{Med}_{i}(\boldsymbol{x};  W)$    and  $\mbox{Med}_{i}(\boldsymbol{x};  W) $ is the  weighted median closet to $x_i$, we have
 $x_{k_a}=x_{k_l}=\mbox{Med}_{i}(\boldsymbol{x}; W)$, which means (\ref{lemma01}) holds by (\ref{pl012b}). This complete the proof of Lemma \ref{lemma0}.
\end{IEEEproof}

The \emph{cohesive set} is an elaborate network structure which was first proposed in  \cite{morris2000contagion}
and specialized in \cite{mei2022convergence}. Following \cite{mei2022convergence}
 we adopt the specific version of the cohesive set.
\begin{definition}[Cohesive set]
	\label{def1}
If a nonempty subset $\mathcal{M}\subset \mathcal{V}$ satisfies $\sum _{j\in \mathcal{M}}w_{ij}\geq 1/2$ for any $i\in \mathcal{M}$, then we say $\mathcal{M}$ is a  \emph{cohesive set } of $\mathcal{G}(W)$.
\end{definition}
We use the following lemma to illustrate the role of cohesive set in opinion update.
\begin{lemma}\label{exle_1}
Consider the opinion dynamics (\ref{sys2}).
If there exists a  cohesive set  $\mathcal{M}\subset \mathcal{V}$ consisting of unprejudiced agents only, then
\begin{equation*}\label{exle_1_1}
	\underset{j\in \mathcal{M}}{\min}x_j(0) \leq x_i(t) \leq \underset{j\in \mathcal{M}}{\max }x_j(0),~~\forall i\in \mathcal{M}, t\in\mathbb{N}.
\end{equation*}
\end{lemma}
\begin{IEEEproof}
Since all agents in $\mathcal{M}$ are unprejudiced,  by
(\ref{sys2}) we have
\begin{equation}\label{exle_1_a1}
x_i(t+1)=\mbox{Med}_{i}(\boldsymbol{x}(t);  W),~~\forall i\in \mathcal{M}, t\in\mathbb{N}.
\end{equation}
Also, since $\mathcal{M}$ is a cohesive set, by Lemma \ref{lemma0}, Definition \ref{def1} and (\ref{exle_1_a1}) we have
\begin{equation}\label{exle_1_2}
\underset{j\in \mathcal{M}}{\min }{x}_{j}(t) \leq x_{i}(t+1) \leq \underset{j\in \mathcal{M}}{\max }{x}_{j}(t),~~ \forall i\in \mathcal{M}, t\in\mathbb{N}.
\end{equation}
From (\ref{exle_1_2}) it is clear that
\begin{equation}\label{exle_1_3}
\underset{j\in \mathcal{M}}{\min }{x}_{j}(t) \leq \min_{i\in\mathcal{M}} x_{i}(t+1)\leq \max_{i\in\mathcal{M}} x_{i}(t+1) \leq \underset{j\in \mathcal{M}}{\max }{x}_{j}(t),~\forall t\in\mathbb{N}.
\end{equation}
Using (\ref{exle_1_3}) repeatedly we can obtain the following inequality
\begin{equation*}\label{exle_1_4}
\underset{j\in \mathcal{M}}{\min }{x}_{j}(0) \leq \min_{i\in\mathcal{M}} x_{i}(t)\leq \max_{i\in\mathcal{M}} x_{i}(t) \leq \underset{j\in \mathcal{M}}{\max }{x}_{j}(0),~\forall t\in\mathbb{N},
\end{equation*}
which means that the lemma is proved.
\end{IEEEproof}

From Lemma \ref{exle_1}, if some unprejudiced  agents form a cohesive set $\mathcal{M}$, then they do not
adopt the opinion beyond the set $\mathcal{M}$ during the opinion evolution. In this sense, a cohesive set can be seen as an echo chamber,  in which
 opinions are disseminated and reinforced inside a closed system \cite{mei2022convergence}. So, to reach consensus, a group should not contain cohesive sets to avoid the generation of echo chambers.

Next we give a necessary and sufficient condition for consensus of the system (\ref{sys2}).
\begin{theorem}[Consensus of weighted-median  opinion dynamics with partial prejudice]
	\label{Main_result1}
The system (\ref{sys2}) achieves asymptotic consensus for any $\boldsymbol{x}(0)\in \mathbb{R}^n$  if and only if  $\mathcal{G}(W)$  does not contain a cohesive set consisting of unprejudiced agents only.
\end{theorem}

\begin{remark}
When $\mathcal{G}(W)$ contains a cohesive set $\mathcal{M}$ consisting of unprejudiced agents only,  the system (\ref{sys2}) may not converge.
For example, consider the system (\ref{sys2}) with $3$ agents, in which agent $1$ is prejudiced and agents $2,3$ are unprejudiced. Let $w_{23}>1/2\mbox{~and~}w_{32}>1/2$.  Then, agents $\{2,3\}$ form a cohesive set. By  (\ref{sys2}), we have  $x_2(t+1)=x_3(t)$ and $x_3(t+1)=x_2(t)$  at each step $t$. The system can not converge to the same opinion.
\end{remark}

\begin{remark}
Consider one or a few prejudiced agents with the same prejudice value joining in an unprejudiced group. The connectivity of the influence network is not enough to ensure that agents reach consensus under the system (\ref{sys2}), but can ensure that agents always synchronize to the unique prejudice value under the FJ model (\ref{FJ}) \cite{proskurnikov2017opinion}. This is the essential difference between the FJ model (\ref{FJ}) and the system (\ref{sys2}). Our system may be more in line with certain social realities that people tend to become similar when they interact, but not all differences will eventually disappear \cite{abelson1964mathematical,axelrod1997dissemination}.

\end{remark}

\subsection{Proof of Theorem \ref{Main_result1}}

Before the proof of Theorem \ref{Main_result1}, we need to introduce some lemmas.
Recall that $\mathcal{V}_1$ and $\mathcal{V}_2$ are the sets of all prejudiced and unprejudiced agents respectively. Let ``$a \vee b $" be the larger of $a,b$ and ``$a \wedge b $" be the smaller of $a,b$.
\begin{lemma}\label{lemma3}
Consider the opinion dynamics (\ref{sys2}).
	If $\mathcal{G}(W)$  does not contain a cohesive set consisting of unprejudiced agents only, then
	\begin{equation}
		\label{gs0}
		\underset{\underset{t-n_2 \leq s \leq t-1}{j\in \mathcal{V}_1}}{\min}x_j(s)
		\leq {x}_{i}(t) \leq
		\underset{\underset{t-n_2 \leq s \leq t-1}{j\in \mathcal{V}_1}}{\max}x_j(s), ~~\forall i \in \mathcal{V}_2, t\geq n_2.
	\end{equation}
	
\end{lemma}
\begin{IEEEproof}
By the condition of this lemma,  $\mathcal{V}_2$  and all its  subset are not cohesive sets. Then from Definition \ref{def1}, there exists an agent $k_1\in\mathcal{V}_2$ such that  $\sum_{j\in\mathcal{V}_2}w_{k_1 j}<1/2$.  Thus we have
\begin{equation}\label{le3_1}
\sum_{j\in \mathcal{V}_1}w_{k_1j}=1- \sum_{j\in \mathcal{V}_2}w_{k_1j}> \dfrac{1}{2}.
\end{equation}
Since $\mathcal{V}_2\setminus \left\{k_1\right\}$ is  not a cohesive set either, there exists an agent $k_2 \in \mathcal{V}_2\setminus \left\{k_1\right\}$ such that $\sum_{j\in \mathcal{V}_2\setminus \left\{k_1\right\}}{w}_{k_2j} <1/2$, which implies
\begin{equation*}\label{le3_2}
{w}_{k_2k_1}+	\sum_{j\in \mathcal{V}_1}w_{k_2j} =1-\sum_{j\in \mathcal{V}_2\setminus \left\{k_1\right\}}{w}_{k_2j} > \dfrac{1}{2}.
\end{equation*}

Repeat the above discussion, we can get ${k_3},  {k_4},  \ldots,  {k_{n_2}}$ such that $k_{i}\in\mathcal{V}_2\setminus\{k_1,\ldots,k_{i-1}\}$ and
\begin{equation} \label{le3_4}
	 \sum_{1\leq l\leq i-1}{w}_{k_ik_l}+\sum_{j\in \mathcal{V}_1}w_{k_ij}> \dfrac{1}{2}  ,~~  \forall i\in\{3,\ldots, n_2\}.
\end{equation}
Combining (\ref{le3_1})-(\ref{le3_4}) with Lemma \ref{lemma0} yields
\begin{eqnarray}\label{gs2}
\begin{aligned}
&~~~~\underset{j\in \mathcal{V}_1}{\min} x_j(t-1) \wedge \underset{1\leq l\leq i-1}{\min} x_{k_l}(t-1) \\
 &\leq  {x}_{k_i}(t)=\mbox{Med}_{k_i}(\boldsymbol{x}(t-1);  W) \\
 &\leq  \underset{j\in \mathcal{V}_1}{\max} x_j(t-1) \vee \underset{1\leq l\leq i-1}{\max} x_{k_l}(t-1),~\forall 1\leq i \leq n_2, t \geq 1.
 \end{aligned}
\end{eqnarray}

	Next we use induction to prove that the inequality
	\begin{equation} \label{le3_5}
			\underset{j\in \mathcal{V}_1}{\max} x_j(t-1) \vee \underset{1\leq l\leq i-1}{\max} x_{k_l}(t-1)\leq 	\underset{\underset{t-i \leq s \leq t-1}{j\in \mathcal{V}_1}}{\max}x_j(s)
	\end{equation}
holds for all $1\leq i \leq n_2$ and $t\geq i$. First,
	it is clear that the inequality (\ref{le3_5}) holds  when  $i=1$ and $t\geq 1$. We assume that the inequality (\ref{le3_5}) holds  for all $i=1,2,\ldots,i^*-1$ and $t\geq i$ with $2\leq i^*\leq n_2$. Then, from this assumption and the second inequality of (\ref{gs2}) it can be directly obtained that
	\begin{multline*}
	{x}_{k_i}(t)  \leq  \underset{\underset{t-i \leq s \leq t-1}{j\in \mathcal{V}_1}}{\max}x_j(s) \leq  \underset{\underset{t-i^*+1 \leq s \leq t-1}{j\in \mathcal{V}_1}}{\max}x_j(s),\\
 ~~\forall 1\leq i \leq i^*-1 , t\geq i^*-1,
	\end{multline*}
which is followed by
	\begin{equation*}\label{le3_6}
	\underset{1\leq i\leq i^*-1}{\max}{x}_{k_i}(t)  \leq   \underset{\underset{t-i^*+1 \leq s \leq t-1}{j\in \mathcal{V}_1}}{\max}x_j(s),~~\forall t\geq i^*-1.
	\end{equation*}
From this we have,
	\begin{align*}
		&\quad 	\underset{j\in \mathcal{V}_1}{\max} x_j(t-1) \vee \underset{1\leq l\leq i^*-1}{\max} x_{k_l}(t-1)& \\
		& \leq \underset{j\in \mathcal{V}_1}{\max} x_j(t-1) \vee \underset{\underset{t-i^* \leq s \leq t-2}{j\in \mathcal{V}_1}}{\max}x_j(s)& \\
		&= \underset{\underset{t-i^* \leq s \leq t-1}{j\in \mathcal{V}_1}}{\max}x_j(s),~~\forall t\geq i^*. &
	\end{align*}
	Up to now, we have proved (\ref{le3_5}) when $i=i^*$ and $t\geq i^*$, so the induction argument is completed.

	Similar to (\ref{le3_5}) we can show that  for any $1\leq i\leq n_2$ and $t\geq i$,
	\begin{align}\label{exq1}
		\underset{j\in \mathcal{V}_1}{\min} x_j(t-1) \wedge \underset{1\leq l\leq i-1}{\min} x_{k_l}(t-1) \geq
		\underset{\underset{t-i \leq s \leq t-1}{j\in \mathcal{V}_1}}{\min}x_j(s)  .
	\end{align}
	By (\ref{gs2}), (\ref{le3_5}) and (\ref{exq1}) we have
	\begin{equation} \label{lem3_10}
		\underset{\underset{t-i \leq s \leq t-1}{j\in \mathcal{V}_1}}{\min}x_j(s)
		\leq {x}_{k_i}(t) \leq
		\underset{\underset{t-i \leq s \leq t-1}{j\in \mathcal{V}_1}}{\max}x_j(s),~~\forall 1\leq i\leq n_2,~t\geq i,
	\end{equation}
	which indicates that
	\begin{multline} \label{lem3_20}
		\underset{\underset{t-n_2 \leq s \leq t-1}{j\in \mathcal{V}_1}}{\min}x_j(s)
		\leq {x}_{k_i}(t) \leq
		\underset{\underset{t-n_2 \leq s \leq t-1}{j\in \mathcal{V}_1}}{\max}x_j(s),\\
\forall 1\leq i\leq n_2, ~t\geq n_2.
	\end{multline}
	Since $\{k_1,k_2,\ldots,k_{n_2}\}=\mathcal{V}_2$,
the result of the lemma can be obtained by (\ref{lem3_20})  immediately.
\end{IEEEproof}

\begin{lemma}\label{lemma2}
	Consider the opinion dynamics (\ref{sys2}).

(i) If there exist $T\geq0$ such that $u\leq\max_{i\in\mathcal{V}} x_i(t)$ for any  $t\geq T$, then $\max_{i\in\mathcal{V}} x_i(t)$ is  monotonically non-increasing for $t\geq T$.

(ii) If there exist $T\geq0$ such that $u\geq\min_{i\in\mathcal{V}} x_i(t)$ for any  $t\geq T$, then $\min_{i\in\mathcal{V}} x_i(t)$ is monotonically  non-decreasing for  $t\geq T$.
\end{lemma}
\begin{IEEEproof}	
(i) For any $i\in\mathcal{V}_1$, by (\ref{sys2}) we have
	\begin{eqnarray*}
\begin{aligned}
		x_i(t+1)&=\lambda_{i}u+(1-\lambda_{i}){\mbox{Med}_{i}(\boldsymbol{x}(t);  W)}\\
		&\leq \lambda_{i} \max_{j\in\mathcal{V}} x_j(t) +(1-\lambda_{i})\max_{j\in\mathcal{V}} x_j(t)\\
&=\max_{j\in\mathcal{V}} x_j(t),~~\forall  t\geq T,
\end{aligned}
\end{eqnarray*}
which means
\begin{align*}
	\max_{i\in \mathcal{V}_1}x_i(t+1) \leq \max_{j\in\mathcal{V}} x_j(t),~~\forall  t\geq T,
	\end{align*}
on the other hand, by (\ref{sys2}) we can obtain that
	\begin{equation}\label{123}
	\max_{i\in \mathcal{V}_2} x_i(t+1)=\max_{i\in \mathcal{V}_2} \mbox{Med}_{i}(\boldsymbol{x}(t);  W) \leq  \max_{j\in\mathcal{V}} x_j(t),~~\forall  t\geq T.
	\end{equation}
From (\ref{123}) we see that $\max_{i\in\mathcal{V}} x_i(t)$ is monotonically non-increasing for  $t\geq T$.

(ii) The proof of (ii) is similar to (i).
\end{IEEEproof}

\begin{lemma}\label{lemma51}
	Consider the opinion dynamics (\ref{sys2}).\\
(i) If there exists $T\geq 0$ such that $u\leq \min_{i\in\mathcal{V}}x_i(T)$, then $u\leq \min_{i\in\mathcal{V}}x_i(t)$ for all $t\geq T$, and $\max_{i\in\mathcal{V}}x_i(t)$ is monotonically non-increasing for $t\geq T$.\\
(ii) If there exists $T\geq 0$ such that $u\geq \max_{i\in\mathcal{V}}x_i(T)$, then  $u\geq \max_{i\in\mathcal{V}}x_i(t)$ for all $t\geq T$, and $\min_{i\in\mathcal{V}}x_i(t)$ is monotonically non-decreasing for $t\geq T$.
\end{lemma}
\begin{IEEEproof}
(i) Since $u\leq \min_{i\in\mathcal{V}}x_i(T)$, by (\ref{sys2}) we have
\begin{eqnarray*}
	\begin{aligned}
		x_i(T+1)&=\lambda_{i}u+(1-\lambda_{i}){\mbox{Med}_{i}(\boldsymbol{x}(T);  W)}\\
		&\geq \lambda_{i}u +(1-\lambda_{i})\min_{j\in\mathcal{V}} x_j(T)\\
		&\geq \lambda_{i}u +(1-\lambda_{i})u=u,~~\forall i\in \mathcal{V}_1,
	\end{aligned}
\end{eqnarray*}
which means
\begin{equation}\label{lemma51_1}
\min_{i\in \mathcal{V}_1}x_i(T+1) \geq u.
\end{equation}
On the other hand, by (\ref{sys2}) we get
\begin{equation}\label{lemma51_2}
	\min_{i\in \mathcal{V}_2} x_i(T+1)=\min_{i\in \mathcal{V}_2} \mbox{Med}_{i}(\boldsymbol{x}(T);  W) \geq  \min_{j\in\mathcal{V}} x_j(T)\geq u.
\end{equation}
Repeat (\ref{lemma51_1}) and (\ref{lemma51_2}) we get
 $u\leq \min_{i\in\mathcal{V}}x_i(t)$ for all $t\geq T$.
By Lemma \ref{lemma2} (i), we obtain that $\max_{i\in\mathcal{V}}x_i(t)$ is monotonically non-increasing for $t\geq T$.

(ii) The proof of (ii) is similar to (i).
\end{IEEEproof}

\begin{lemma}\label{lemma4}
	Consider the opinion dynamics (\ref{sys2}). We assume that $\mathcal{G}(W)$  does not contain a cohesive set consisting of unprejudiced agents only.\\
(i) If there exists $T\geq 0$ such that  $\max_{i\in\mathcal{V}}x_i(t)$ is monotonically non-increasing for $t\geq T$,
 then
	\begin{multline}
		\label{gs01}
		 {x}_{i}(t)-u \leq (1-\lambda_{\min})^K \Big(\max_{j\in\mathcal{V}} x_j(T)-u\Big),\\
\forall i\in \mathcal{V}_1, K\in\mathbb{Z}^+, t\geq (K-1)(n_2+1)+T+1.
	\end{multline}

(ii) If there exists $T\geq 0$ such that $\min_{i\in\mathcal{V}}x_i(t)$ is monotonically non-decreasing for $t\geq T$, then
	\begin{multline}
		\label{gs02}
		{x}_{i}(t)-u \geq (1-\lambda_{\max})^K\Big(\min_{j\in\mathcal{V}} x_j(T)-u \Big),\\
\forall i\in \mathcal{V}_1, K\in\mathbb{Z}^+, t\geq (K-1)(n_2+1)+T+1.
	\end{multline}
\end{lemma}

\begin{IEEEproof}
(i)	We use induction to prove (\ref{gs01}).
	
	When $K=1$,  by  (\ref{sys2}) we have
	\begin{align*}
		{x}_{i}(t)-u&=(1-\lambda_{i})\big(\mbox{Med}_{i}(\boldsymbol{x}(t-1);W)-u\big)
\\ &\leq (1-\lambda_{\min})\Big(\max_{j\in\mathcal{V}} x_j(t-1) -u\Big)
\\ &\leq (1-\lambda_{\min})\Big(\max_{j\in\mathcal{V}}x_j(T)-u \Big),~\forall i\in \mathcal{V}_1, t\geq T+1.
	\end{align*}
	So (\ref{gs01}) holds for  $K=1$.

	Assume (\ref{gs01}) holds for $K \leq L$.
	By Lemma \ref{lemma3} we have
	\begin{align}\label{le5_ex1}
		 \max_{j \in \mathcal{V}_2} x_{j}(t)\leq \underset{\underset{t-n_2 \leq s \leq t-1}{k \in \mathcal{V}_1}}{\max}x_k(s),~~ t\geq n_2.
	\end{align}
	Then, by (\ref{le5_ex1}) we get
    \begin{align}\label{le5_ex2}
    	&\max_{j \in \mathcal{V}} x_{j}(t)= \underset{j \in \mathcal{V}_1}{\max}{x}_{j}(t) \vee \underset{j \in \mathcal{V}_2}{\max}{x}_{j}(t)&\\
     &~~~~~~~~~~~~~\leq  \underset{\underset{t-n_2 \leq s \leq t}{k \in \mathcal{V}_1}}{\max}x_k(s),~~~~~~~~~~ \forall t\geq n_2.&
    \end{align}

  For any $t\geq L(n_2+1)+T$, let  $i_t\in\mathcal{V}_1$ and  ${t}^{\ast}\in \{t-n_2,\ldots,t\}$ satisfy
    \begin{align*}
    	{x}_{{i}_{t}}({t}^{\ast})=\underset{\underset{t-n_2 \leq s \leq t}{k \in \mathcal{V}_1}}{\max}x_k(s).
    \end{align*}
    Since ${t}^{\ast}\geq t-n_2 \geq (L-1)(n_2+1)+T+1$ and (\ref{gs01}) holds for $K=L$,  by (\ref{le5_ex2}) we have
    \begin{multline}\label{le5_ex3}
    	\max_{j\in\mathcal{V}}x_j(t) \leq {x}_{{i}_{t}}({t}^{\ast})\leq
    	(1-\lambda_{\min})^L\Big(\max_{j\in\mathcal{V}}x_j(0)-u\Big)+u,\\~~\forall t\geq L(n_2+1)+T.
    \end{multline}
    By  (\ref{sys2}) and (\ref{le5_ex3}) we have
    \begin{align*}
    	&~~~~{x}_{i}(t)-u\\&=(1-\lambda_i)\big(\mbox{Med}_{i}(\boldsymbol{x}(t-1);W)-u\big)&\\&\leq (1-\lambda_{\min})\Big(\max_{j\in\mathcal{V}} x_j(t-1)-u\Big)&\\
    &	\leq (1-\lambda_{\min})\Big((1-\lambda_{\min})^L\big(\max_{j\in\mathcal{V}} x_j(0)-u\big)+u-u\Big)&\\
    	&=(1-\lambda_{\min})^{L+1}\Big(\max_{j\in\mathcal{V}} x_j(0)-u\Big),\\
    &~~~~~~~~~~~~~~~~~~~~~~~~~~~~~~~~~~~~~~~ \forall i\in \mathcal{V}_1,t\geq L(n_2+1)+T+1. &
    \end{align*}

    Up to now, we have proved (\ref{gs01}) holds when $K=L+1$. By induction, we see that (\ref{gs01}) holds for all $K\in\mathbb{Z}^+$.

    (ii)  The proof of (ii) is similar to (i).
\end{IEEEproof}

Next we  give the proof of Theorem \ref{Main_result1}.
\begin{IEEEproof}[Proof of Theorem \ref{Main_result1}]
$(\Longleftarrow)$
Consider the  following two cases:

Case I:  The  state of system (\ref{sys2}) satisfies $ \min_{i\in\mathcal{V}} x_i(t)<u<\max_{i\in\mathcal{V}} x_i(t)$ for any $t\geq 0$.
From Lemma \ref{lemma2}, we get  $\max_{i\in\mathcal{V}} x_i(t)$ and  $\min_{i\in\mathcal{V}} x_i(t)$ are  monotonically non-increasing and non-decreasing  respectively for $t\geq 0$.
Because $\lambda_{\min}\in(0,1]$, by Lemma \ref{lemma4} we have  \[ \underset{t\to \infty }{\lim}{x}_{i}(t)=u,~ i \in \mathcal{V}_1 .\]

Case II: There exists  $T\geq0$ such that $u\leq \min \boldsymbol{x}(T)$ or $u\geq \max \boldsymbol{x}(T)$.
 Since the analysis of these two cases are similar, we assume $u\leq \min \boldsymbol{x}(T)$ without loss of generality.
By Lemma \ref{lemma51} (i), we have $$x_i(t)-u\geq 0,~~\forall i\in\mathcal{V}_1,t\geq T,$$
and $\max_{i\in\mathcal{V}}x_i(t)$ is monotonically non-increasing for $t\geq T$.
Combining this with Lemma \ref{lemma4} (i), for all prejudiced agents we have  \[ \underset{t\to \infty }{\lim}{x}_{i}(t)=u,~~\forall i \in \mathcal{V}_1. \]

Together Cases I and II with Lemma \ref{lemma3}, for any $\boldsymbol{x}(0)\in\mathbb{R}^n$ we have \[ \underset{t\to \infty }{\lim}{x}_{i}(t)=u,~~\forall i \in \mathcal{V}_1\cup \mathcal{V}_2. \]

$(\Longrightarrow)$
We give the proof for this result by contradiction.
Let $\mathcal{M}\subset \mathcal{V}$ be a cohesive set composed by  unprejudiced agents only.
 By Lemma \ref{exle_1}, for all $i\in \mathcal{M}$ and $t\geq 0$,  we have
  \begin{equation}
  	\label{15}
 	\underset{j\in \mathcal{M}}{\min }{x}_{j}(0) \leq {x}_{i}(t) \leq \underset{i\in \mathcal{M}}{\max }{x}_{j}(0).
 \end{equation}
Let $a$ be a real number satisfying $a>u$. Choose $x_i(0)=a$ for any $i\in\mathcal{V}$.
Since $\mathcal{M}\subset \mathcal{V}_2$,  from (\ref{15}) we have
 \begin{equation}\label{15-new}
  {x}_{i}(t)=a,~~~~ \forall i\in \mathcal{M},  t\geq 1.
 \end{equation}
 Then by (\ref{sys2}), for any $i\in \mathcal{V}_1$ we have
 \begin{align*}
 	{x}_{i}(1)&=\lambda_iu+(1-\lambda_i)\mbox{Med}_{i}(\boldsymbol{x}(0);  W)& \\ &\leq
 	\lambda_iu+(1-\lambda_i)a=a-\lambda_i(a-u)<a.
 \end{align*}
 Similarly, we can get ${x}_{i}(t)\leq a-\lambda_i(a-u)$ for any $i\in \mathcal{V}_1$ and $t=2,3,\ldots$. By this and (\ref{15-new}) we see that the consensus cannot be reached. This complete the proof of the theorem.
\end{IEEEproof}

\renewcommand{\thesection}{\Roman{section}}
\section{Simulations}\label{NE3}
\begin{figure}
	\centering
	\includegraphics[width=2.5in]{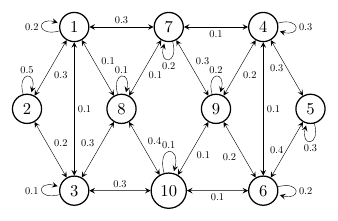}
	\caption{The symmetrical influence network $\mathcal{G}(W)$ used in our simulations.
}\label{Figb}
\end{figure}

\renewcommand{\thesection}{\arabic{section}}
 Consider a symmetric influence network   $\mathcal{G}(W)$ shown in Fig. \ref{Figb}  with $n=10$ agents.
 The initial state $\boldsymbol{x}(0)$ of agents is set to be $ (0.5,0.4,0.3,0.2,0.1,0,-0.1,-0.2,-0.3,-0.4)^{\top}$. For each prejudiced agent $i\in\mathcal{V}_1$, we set its prejudice $u_i=x_i(0)$ and choose  its susceptibility $\lambda_i$ randomly and uniformly from  $(0,1]$. With this same configuration  we compare the opinion evolution between the weighted-median mechanism (\ref{doc10}) and FJ model (\ref{FJ}).

\begin{figure}
	\centering
	\subfigure[]{
		\label{Fig.sub.a}
		\includegraphics[width=1.6in]{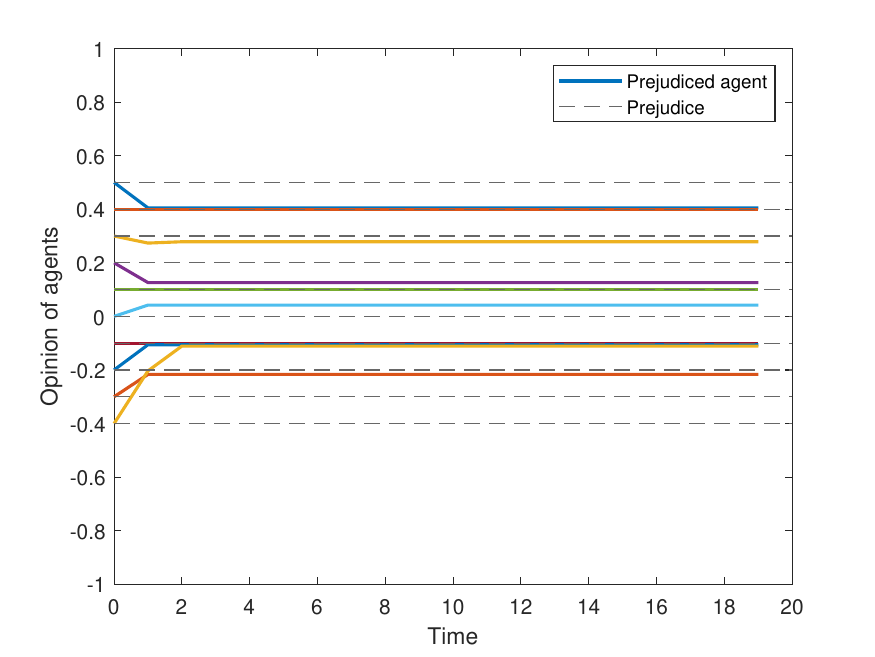}}
	\subfigure[]{
		\label{Fig.sub.b}
		\includegraphics[width=1.6in]{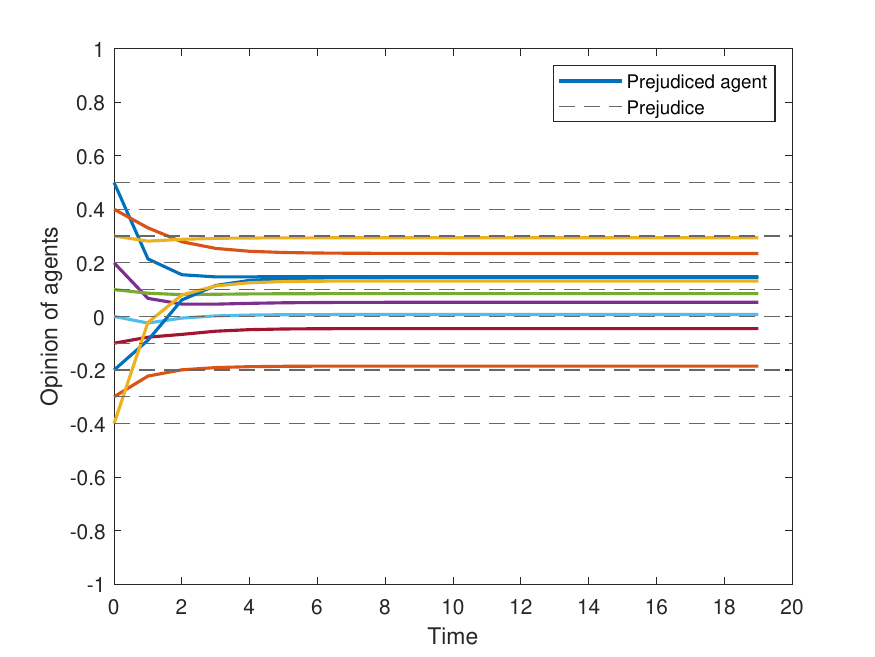}}
\subfigure[]{
		\label{Fig.sub.c}
		\includegraphics[width=1.6in]{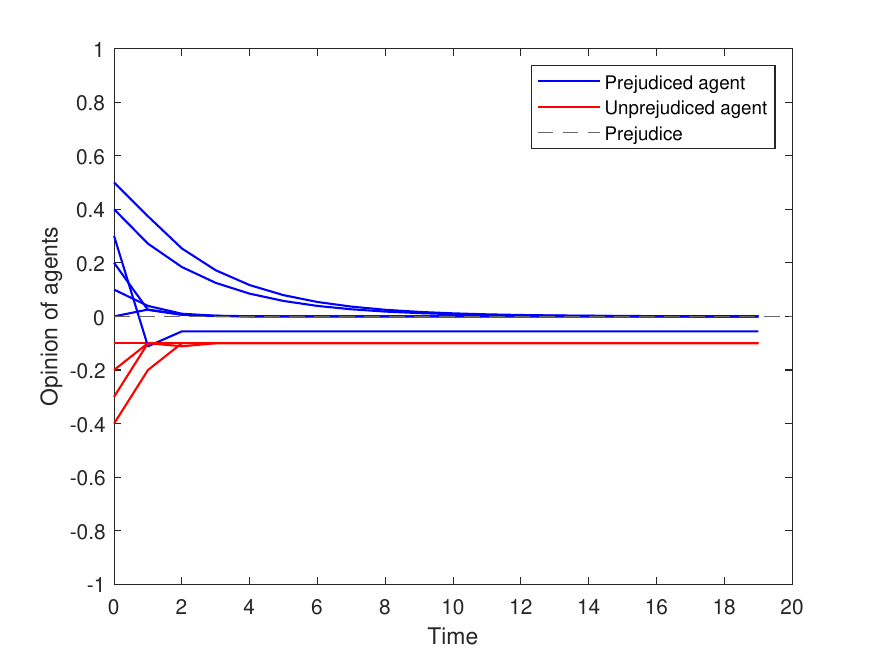}}
	\subfigure[]{
		\label{Fig.sub.d}
		\includegraphics[width=1.6in]{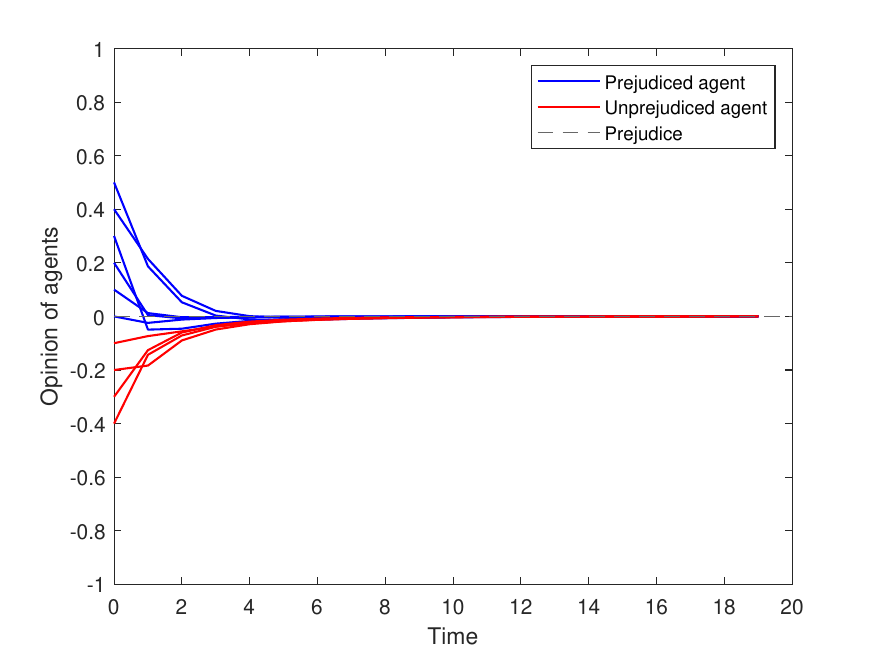}}
	\caption{Opinion evolutions of our system (\ref{doc10}) with prejudiced agents (a) and  mixed agents (c),
 and the FJ model (\ref{FJ}) with prejudiced agents (b) and mixed agents (d).}
	\label{Fig.maina}
\end{figure}

When all agents are prejudiced, the opinion evolutions of our system (\ref{doc10}) and the FJ model (\ref{FJ}) are shown in Figures \ref{Fig.sub.a} and \ref{Fig.sub.b} respectively.
It can be observed that these two systems exhibit similar convergence behaviors in this case.
 When agents $\{1,\ldots,6\}$ are prejudiced agents and $\{7,\ldots,10\}$ are unprejudiced agents,  the opinion evolutions of these two systems (\ref{doc10}) and (\ref{FJ}) are shown in Figures \ref{Fig.sub.c} and  \ref{Fig.sub.d} respectively. It can be observed that the system  (\ref{doc10}) is essentially different from the FJ model  (\ref{FJ}) in this case. The former converges to multiple clusters, while the latter converges to consensus.

\renewcommand{\thesection}{\Roman{section}}
\section{Concluding Remarks}\label{CR}

This paper gives a theoretical analysis of the dynamical behavior of the weighted-median opinion dynamics with prejudice. First, we provide the convergence and the  negative
 exponential convergence rate when all agents are prejudiced. Also, the explicit expression of the limit point is given in form. Moreover,
 with the concept of cohesive set, we characterize  the sufficient and necessary conditions for our system to achieve consensus asymptotically when agents are mixed.

  It remains some problems to address in the future. For example, the system (\ref{sys2}) only considers the case when all prejudiced agents have the same prejudice. However, if they have different prejudices the convergence of  system (\ref{sys2}) is still unknown. Another interesting problem is to find the minimal set of prejudiced agents leading all unprejudiced  agents to reach consensus.
According to Theorem \ref{Main_result1}, this problem is to find a minimal prejudiced set such that the influence network does not contain an unprejudiced  cohesive set.
We leave them for future work.



\renewcommand{\thesection}{\arabic{section}}

%

\ifCLASSOPTIONcaptionsoff
  \newpage
\fi

\bibliographystyle{IEEEtran}
\bibliography{ckwx}

\end{document}